\documentclass{iau} 
\usepackage{graphicx}

\title[Kinematics and Dynamics of Young Star Clusters with TMT] 
{Kinematics and Dynamics of Young Star Clusters with TMT}

\author[Priya Shah]   
{Priya Shah}

\affiliation{Department of Physics, Maulana Azad National Urdu University,\\ 
Gachibowli, Hyderabad 500 032, India \\ email: {\tt priya.hasan@gmail.com} }

\pubyear{2018}
\volume{347}  
\jname{Early Science with ELTs (EASE)}
\editors{Norbert Przybilla, Karen Pollard \& Annalisa Calamida, eds.}
\begin{document}

\maketitle

\begin{abstract}
Star clusters are evolving n-body systems. We discuss the early dynamics of star clusters, the process of primordial mass segregation and clustering observed in young clusters.  We discuss how dynamics coupled with stellar evolution in a cluster define the radial profile, mass function and disruption of the cluster and compare these parameters with some known clusters. As a member of the Thirty Meter Telescope (TMT), International Science Driven Team (ISDT),  I shall use these details to help define the science case, requirements  and the expected precision in answering possible questions about the evolution of star clusters in terms of astrometry and high resolution spectroscopy. I shall also report on some of the resolutions made at the recent TMT Forum held in Mysore, India. 
\keywords{open clusters and associations: general, galaxies: star clusters, stars: formation, stellar dynamics, telescopes, astrometry}
\end{abstract}

\firstsection 
\section{Introduction}
Star clusters are the smallest organised blocks of stars which embody information about their formation, evolution and dynamics. They exist in a variety of manifestations like associations, embedded systems, open systems and dense globular clusters. Understanding the relation between their structure, distribution and velocities of young stars and their dynamical evolution are key elements to our understanding of the star formation process. 

The present understanding is that stars form in groups or aggregates and eventually due to gas dispersal and/or gravitational interactions, most of them disperse in the field and only a few remain gravitationally bound (\cite[Lada \& Lada 2003]{2003ARA&A..41...57L}, \cite[Gutermuth et al. 2009]{gut09}, \cite[Gouliermis 2018]{2018PASP..130g2001G}). In very young clusters, the Young Stellar Objects (YSOs) are embedded or partially embedded in star-forming clouds and typically have clumpy distributions. A kinematic study of them will help answer big questions about cluster assembly, equilibration and dissolution which will help draw conclusions on theories of monolithic and hierarchical formation of clusters.

Mass segregation is the distribution of stars in a cluster according to their masses, i.e., where the more massive stars are closer to the center than the low mass stars. Primordial or initial mass segregation affects the dynamical evolution of clusters. In an earlier paper, \cite[Hasan \& Hasan (2011)]{Hasbro} studied mass segregation in a sample of nine clusters from the Two~Micron~All~Sky~Survey (2MASS) (\cite[Skrutskie et al. 2006]{skru06}). The authors could not make conclusive statements whether mass segregation is a birth phenomenon, or whether the more massive stars form anywhere throughout the proto-cluster volume and then migrate due to gravitational dynamical effects.  
Assuming primordial mass segregation would imply that massive stars only form in rich clusters and reject the possibility they can also form in isolation (\cite[Parker \& Goodwin 2007]{par07}). Detailed simulations have been presented in \cite[Vesperini et al. (2009)]{verp09}.

In most studies of young clusters, there is always an issue of obtaining reliable samples of cluster members in nebulous regions with field star contaminantion.  The absence of kinematic information for faint stars causes such studies to be incomplete and indecisive.  

A recent paper by \cite[Kuhn et al.(2018)]{2018arXiv180702115K}  studies YSOs using the Gaia DR2 data, in 28 clusters and associations at distances ranging from 0.3-3.7 kpc\footnote{A 1 mas/yr measurement will imply a tangential velocity of $\approx$ 1 km/s at a distance of 4 kpc, $\mu(arc sec/yr)=\frac{V_t(km/s)}{4.74 d(pc)}$}. They found that  70\% of these clusters show signs of expansion, with typical expansion velocities of  $\approx$ 0.5 km/s and a positive radial gradient in the  expansion velocity (positional sorting). Some clusters like Orion Nebula Cluster, NGC 6231, NGC 2362 showed no expansion. The velocity dispersion measured were in the range  1-3 km/s.

The limitation of Gaia is that it observes in the optical and most young stars are deeply embedded in dust and gas. Gaia has issues with extremely red objects, extincted regions and brown dwarfs. The kinematics within individual sub-clusters is not complete,  due to insufficient numbers of stars in the sample used. Hence, we require complementary data for  more accurate radial velocities and proper motions at fainter magnitudes and in extincted regions. 

\section{The Thirty Meter Telescope (TMT)}
 The largest operating optical telescopes in present times  are the Gran Telescopio Canarias(10.4 m), the Large Binocular Telescope (2 $\times$ 8.4 m), Gemini (2 $\times$ 8.1 m), Keck (2 $\times$ 10 m) and the Very Large Telescope (4 $\times$ 8.2 m).  
 
  The large telescopes planned for the next decade are the European Extremely Large Telescope (39.8 m), the Thirty Meter Telescope (TMT, 30 m) and the Giant Magellan Telescope (24.5 m).  The TMT will have a primary mirror of 30 m made up of 492  mirror segments each with a diameter of 1.45 m, operational from 0.31 to 28 $\mu$m, with adaptive optics. This is being planned at Mauna Kea, Hawaii by an international partnership between Caltech, Universities of California, Canada, Japan, China and India (\footnote{http://www.tmt.org/}).  
  
 The 10\% Indian contribution of 13 billion Indian National Rupess (INR) (150 million Euro) will be 70\% in kind and 30\% in cash.  India will have  an opportunity to develop its expertise in telescope making technology, which can be used in development of its own telescopes and also to provide opportunities for the country in this narrow area of expertise. At present, India is making the Edge Sensors, Actuators, Segment Support Assemblies, Segment Polishing and  the  Software Development for the Observatory Software (OSW), Data Management System (DMS) Image and Object Catalogues (CAT)\footnote{http://tmt.iiap.res.in/}.

The International Science Driven Teams (ISDTs) are  working to plan Detailed Science Cases for the telescope with First Light Instruments and Future Planned Instruments including one on  Formation of Stars and Planets of which the author is a member from inception. The Detailed Science Case (\cite[Skidmore et al. 2015]{2015RAA....15.1945S}) describes `the transformational science that the Thirty Meter Telescope will enable'. 

\begin{figure}[h]
\begin{center}
 \includegraphics[width=3.4in]{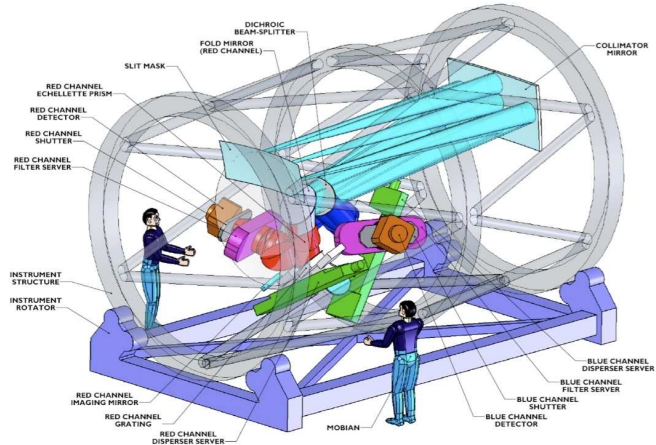} 
 \caption{The Wide Field Optical Spectrograph (WFOS). 
 Courtesy TMT International Observatory. }
   \label{fig1}
\end{center}
\end{figure}

\section{First Light Instruments: The Wide Field Optical Spectrograph (WFOS) and Infrared Imaging Spectrograph (IRIS)}

WFOS is a First Light Instrument and will provide near-ultraviolet and optical (0.31 – 1.0 $\mu$m) imaging and spectroscopy over a 20 to 25 square arc minute field of view. Using precision cut focal plane masks, WFOS will enable short-slit observations of $\approx$ 50 to 60 objects simultaneously. A down selection process is being followed to identify either a slit-mask based instrument concept or a fibre based instrument concept for continued development of the WFOS (Fig. \ref{fig1}). At the TMT forum held recently in India, both options were considered and debated.

\begin{figure}[h]
\begin{center}
 \includegraphics[width=3.4in]{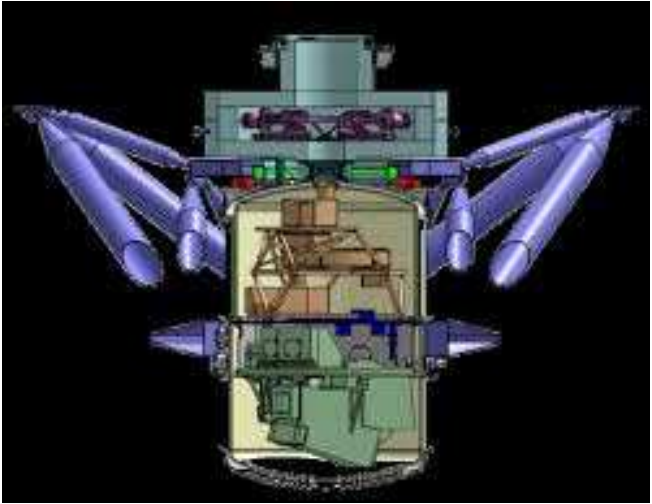} 
 \caption{The Infrared Imaging Spectrograph (IRIS). Courtesy TMT International Observatory. }
   \label{fig2}
\end{center}
\end{figure}

 IRIS (\cite[Larkin et al. 2016]{2016SPIE.9908E..1WL})  is a first generation near-infrared (0.85-2.5 $\mu$m) instrument being designed to sample the diffraction limit of the TMT. IRIS will include an integral field spectrograph (R $\approx$ 4000) and imaging camera ($17''\times 17''$). Both the spectrograph and imager will take advantage of the high spatial resolution achieved with the Narrow-Field Infrared Adaptive Optics System (NFIRAOS) at four spatial scales ($0.004'', 0.009'', 0.025'', 0.05''$). IRIS will achieve an angular resolution ten times better than images from the Hubble Space Telescope, and will be the highest angular resolution near-infrared instrument in the world (Fig. \ref{fig2}).  This is the instrument that shall be put to use for our purpose.

\section{Conclusions}
Young stars in star-forming regions interact with each other in complex ways. Simulations of star cluster evolution generally assume that the stars are initially smoothly distributed and in dynamical equilibrium. However, both observations and theories tell us that this may not be how clusters form.

To answer the question of the nature of early dynamical evolution and star formation, precise mass, position and velocity measurements are required. It is also critical to systematically study the most likely initial conditions for cluster formation which lead to the observed configurations. The radial density profiles of young and embedded clusters can shed light on the structure of those clusters. Colours can help classify types of young stellar objects to trace the progress of star formation and help distinguish between different star formation models. The TMT instruments, IRIS, WFOS and in particular Mid-Infrared Camera High-disperser \& IFU spectrograph (MICHI) will play a very important role on this subject with the very high angular resolution ($\approx$ 10 AU at 140 pc at 10 $\mu$m) and the capability of high-dispersion (R $\approx$ 100,000) spectroscopy. 

This work will answer crucial questions like how does star formation occur, proceed  and triggered?   Is the spatial mass-segregation built-in at birth or is it a result of dynamical evolution? The motions of sub-clusters and clusters of young stars can be observed and studied. A host of questions regarding star formation wait to be answered with the use of this powerful instrument.

\begin{discussion}

\discuss{Jan Palous}{With the expansion velocities and the velocity dispersion of individual clusters, can you constrain the stellar/total masses?}

\discuss{Shah}{Assuming the cluster is virialised, we can derive the masses, which will be an upper bound  and obviously incorrect. The expansion velocities can be used to estimate the departure from virialisation and this seems an interesting problem to work on. }

\end{discussion}


\begin{thebibliography}{}
\bibitem[Gaia Collaboration et al. 2016]{2016A&A...595A...1G} Gaia Collaboration et al. 2016, \textit{A\& A}, 595, A1 
\bibitem[Gouliermis (2018)]{2018PASP..130g2001G} Gouliermis, D.~A.\ 2018, \textit{PASP}, 130, 072001 
\bibitem[Gutermuth et al, 2009]{gut09}
Gutermuth, R. A., Megeath, S. T., Myers, P. C., et al. 2009,
\textit{ApJS}, 184, 18
\bibitem[Hasan \& Hasan 2011]{has211} Hasan, P., \& Hasan,  S. N., 2011, \textit{MNRAS}, 413, 4, 2345  
\bibitem[Kuhn et al.(2018)]{2018arXiv180702115K} Kuhn, M.~A., Hillenbrand, L.~A., Sills, A., Feigelson, E.~D., \& Getman, K.~V.\ 2018, arXiv:1807.02115 
\bibitem[Lada \& Lada(2003)]{2003ARA&A..41...57L} Lada, C.~J., \& Lada, E.~A.\ 2003, 
\textit {ARAA}, 41, 57 
\bibitem[Larkin et al. 2016]{2016SPIE.9908E..1WL} Larkin, J.~E., Moore, A.~M., Wright, S.~A., et al.\ 2016, \textit{Proceedings of SPIE}, 9908, 99081W
\bibitem[Parker \& Goodwin 2007]{par07} Parker R.~J. and Goodwin, S.~P. 2007, \textit {MNRAS}, 380, 1271
\bibitem[Skidmore et al. 2015]{2015RAA....15.1945S} Skidmore, W., TMT International Science Development Teams, \& Science Advisory Committee, T.\ 2015, 
\textit{RAA}, 15, 1945 
\bibitem[Skrutskie et al. 2006]{skru06} Skrutskie, M.F. et al., 2006,  
\textit{\it AJ}, 131, 1163
\bibitem[Vesperini et al. 2009]{verp09}Vesperini E., Stephen L. W., McMillan, Portegies Zwart, S.~F. 2009, \textit{ApJ}, 698, 615

\end{thebibliography}
\end{document}